\input harvmac

\lref\jonghe{F.\ De Jonghe, K.\ Peeters and K.\ Sfetsos,
{\it Killing-Yano supersymmetry in String Theory}, Class.\ Quant.\ Grav.\
{\bf 14} (1997) 35; hep-th/9607203.}
\lref\witsusy{E.\ Witten, {\it Constraints on Supersymmetry Breaking},
Nucl.\ Phys.\ {\bf B202} (1982) 253\skipthis{--316}.}
\lref\jmfd{J.\ Michelson, {\it Scattering of Four-Dimensional Black Holes},
Phys.\ Rev.\ {\bf D 57} (1998) 1092\skipthis{--1097}; hep-th/9708091.}
\lref\dkjm{D.\ M.\ Kaplan and J.\ Michelson, {\it Scattering of Several
Multiply Charged Extremal $D=5$ Black Holes}, Phys.\ Lett.\ {\bf B410} (1997)
125\skipthis{--130}; hep-th/9707021.}
\lref\asad{A.\ Strominger, {\it $AdS_2$ Quantum Gravity and String Theory},
hep-th/9809027.}
\lref\cp{R.\ Coles and G.\ Papadopoulos, {\it The Geometry of the
One-Dimensional Supersymmetric Non-Linear Sigma Models}, Class.\ Quant.\
Grav.\ {\bf 7} (1990) 427.}
\lref\jm{J.\ Maldacena,
{\it The Large $N$ Limit of Superconformal Field
Theories and Supergravity},
Adv. Theor. Math. Phys. {\bf 2} (1998) 231\skipthis{--252};
hep-th/9711200.}
\lref\gaunt{J.\ P.\ Gauntlett, {\it Low-Energy Dynamics of
Supersymmetric Solitons}, Nucl.\ Phys.\ {\bf B400} (1993) 103; hep-th/9205008.}
\lref\pss{ S.\ Paban, S.\ Sethi and M.\ Stern,
{\it Constraints From Extended Supersymmetry in Quantum Mechanics},
Nucl.\ Phys.\ {\bf B534} (1998) 137--154; hep-th/9805018.}
\lref\obata{M.\ Obata, {\it Affine Connections on Manifolds with Almost
Complex, Quaternion or Hermitian Structure}, Japan.\ J.\ Math.\ {\bf 26} 43.}
\lref\intya{K.\ Yano and M.\ Ako, {\it Integrability Conditions for Almost
Quaternion Structures,} Hokkaido Math.\ J.\ {\bf 1} (1972) 63.}
\lref\swt{A.\ Strominger, {\it Superstrings with Torsion},
Nucl.\ Phys.\ {\bf 274} (1986) 253.}
\lref\dps{M.\ Douglas, J.\ Polchinski and A.\ Strominger,
{\it Probing Five-Dimensional Black Holes with $D$-branes},
JHEP {\bf 12} (1997) 003; hep-th/9703031.}
\lref\ghr{S.\ J.\ Gates, Jr., C.\ M.\ Hull and M.\ Ro\v{c}ek,
{\it Twisted Multiplets and New Supersymmetric Non-linear $\sigma$-Models},
Nucl.\ Phys.\ {\bf B248} (1984) 157\skipthis{--186}}.
\lref\gps{G.\ W.\ Gibbons, G.\ Papadopoulos and K.\ S.\ Stelle,
{\it HKT and OKT Geometries on Soliton Black Hole Moduli Spaces}, Nucl.\
Phys.\ {\bf B508} (1997) 623; hep-th/9706207.}
\lref\aipt{J.\ A.\ de Azc\'{a}rraga, J.\ M.\  Izquerido, J.\ C.\ P\'{e}rez Buono
and
P.\ K.\ Townsend, {\it Superconformal Mechanics, Black Holes, and Non-linear
Realizations}, Phys.\ Rev.\ {\bf D59} (1999) 084015; hep-th/9810230.}
\lref\gk{G.\ W.\ Gibbons and R.\ Kallosh, Phys. Rev {\bf D51} (1995) 2839.}
\lref\aff{V.\ de Alfaro, S.\ Fubini and G.\ Furlan, {\it Conformal Invariance
in Quantum Mechanics}, Nuovo Cimento {\bf 34A}
(1976) 569.}
\lref\ap{V.\ P.\ Akulov and I.\ A.\ Pashnev, {\it Quantum Superconformal Model
in (1,2) Space}, Theor.\ Math.\ Phys.\
{\bf 56} (1983) 862.}
\lref\fr{S.\ Fubini and E.\ Rabinovici, {\it Superconformal Quantum
Mechanics}, Nucl.\ Phys.\ {\bf B245} (1984) 17.}
\lref\ikl{E.\ Ivanov, S.\ Krivonos and V.\ Leviant, {\it Geometry of Conformal
Mechanics}, J.\ Phys.\ {\bf A 22},
(1989) 345\semi E. Ivanov, S. Krivonos and V. Leviant,
{\it Geometric Superfield Approach to Superconformal
Mechanics}, J.\ Phys.\ {\bf A 22} (1989) 4201.}
\lref\frme{D. Z. Freedman and P. Mende, {\it An Exactly Solvable $N$ Particle
System in Supersymmetric Quantum Mechanics}, Nucl. Phys. {\bf B344} (1990) 317.}
\lref\cdkk{P. Claus, M. Derix, R. Kallosh, J. Kumar, P. Townsend
and A. van Proeyen, {\it Black Holes and Superconformal Mechanics}, Phys.\
Rev.\ Lett.\ {\bf 81} (1998) 4553; hep-th/9804177.}
\lref\jrmy{J. Michelson and A. Strominger, {\it Superconformal Multi-Black
Hole Quantum Mechanics}, JHEP {\bf 009} (1999) 005; hep-th/9908044.}
\lref\hp{P. Howe and G. Papadopoulos, {\it Further Remarks on the Geometry
of Two-Dimensional Non-Linear $\sigma$-Models}, Comm. Math. Phys. {\bf 151} (1993) 467.}
\lref\jackiw{C.\ G.\ Callan, S.\ Coleman and R.\ Jackiw, {\it A New
Improved Energy-Momentum Tensor}, Ann. Phys. (NY)
{\bf 59} (1970) 42\semi 
R.\ Jackiw, {\it Introducing Scale Symmetry},
Physics Today {\bf 25} (1972) 23.}
\lref\hag{C.\ R.\ Hagan, {\it Scale and Conformal Transformations in
Galilean-Covariant Field Theory}, Phys.\ Rev.\ {\bf D5} (1972)
377\skipthis{--388}.}
\lref\nied{U.\ Niederer, {\it The Maximal Kinematical Invariance Group of the Free
Schr\"{o}dinger Equation}, Helv.\ Phys.\ Acta {\bf 45} (1972)
802\skipthis{--810}.}
\lref\gs{J.-G.\ Zhou, {\it Super 0-brane and GS Superstring Actions on
$AdS_2 \times S^2$}, Nucl.\ Phys.\ {\bf B559} (1999) 92; hep-th/9906013.}
\lref\dewit{B.\ de Wit, B.\ Kleijn and S.\ Vandoren, {\it Rigid $N=2$
Superconformal Hypermultiplets}, in J.\ Wess and E. A. Ivanov (eds.) {\it
Dubna 1997, Supersymmetries and Quantum Mechanics}, Springer, 1999, pp.\
37--45; hep-th/9808160.}

\skip0=\baselineskip
\divide\skip0 by 2
\def\tmpsp{\the\skip0}

\let\linesp=\mylinesp
\def\smaleq{&\!\!\!=&\!\!\!}

\def\nother{Noether}

\def\skipthis#1{{}}

\def\subsubsec#1{\par\noindent {\it #1}\par}

\def\IR{\relax{\rm I\kern-.18em R}}
\def\IZ{\relax\ifmmode\hbox{Z\kern-.4em Z}\else{Z\kern-.4em Z}\fi}
\def\IQ{\relax{\rm I\kern-.40em Q}}
\def\IS{\relax{\rm I\kern-.18em S}}
\def\p{\partial}
\def\slr{{SL(2,\IR)}}

\def\e{\epsilon}
\def\ol#1{{\cal O}(\lambda^{#1})}
\def\anti#1#2{{\{{#1},{#2}\}}}
\def\com#1#2{{[{#1},{#2}]}}

\def\appq{A}
\def\appwhkt{C}

\Title{\vbox{\baselineskip12pt\hbox{hep-th/9907191}
\hbox{HUTP-99/A045}}}{The Geometry of (Super) Conformal Quantum Mechanics}

\centerline{Jeremy Michelson\foot{Current address: New High
Energy Theory Center; Rutgers University; 126 Frelinghuysen Road;
Piscataway, NJ \ 08854.}$^{b,a}$ and Andrew  Strominger$^a$}
\bigskip\centerline{$^a$Department of Physics}
\centerline{Harvard University}
\centerline{Cambridge, MA 02138}

\bigskip\centerline{$^b$Department of Physics}
\centerline{University of California}
\centerline{Santa Barbara, CA  93106}

\vskip .3in \centerline{\bf Abstract} $N$-particle quantum
mechanics described by a sigma model with an $N$-dimensional
target space with torsion is considered. It is shown that an
$\slr$ conformal symmetry exists if and only if the geometry
admits a homothetic Killing vector $D^a\p_a$ whose associated
one-form $D_adX^a$ is closed. Further, the $\slr$ can always be
extended to $Osp(1|2)$ superconformal symmetry, with a suitable
choice of torsion, by the addition of $N$ real fermions. Extension
to $SU(1,1|1)$ requires a complex structure $I$ and a holomorphic
$U(1)$ isometry $D^aI_a{^b}\p_b$. Conditions for extension to the
superconformal group $D(2,1;\alpha)$, which involve a triplet of
complex structures and $SU(2)\times SU(2)$ isometries, are
derived. Examples are given.
\smallskip
\Date{}
\newsec{Introduction}

 Conformal and superconformal field theories in various
dimensions have played a central role in our understanding
of modern field theory and
string theory. Oddly, the subject of this
paper---one dimension---is one of the least well understood cases.
The simplest example of
conformally invariant single-particle
quantum mechanics was pioneered in \aff, following the general analysis
of~\refs{\jackiw\hag{--}\nied}. Supersymmetric
generalizations were discussed in \refs{\ap\fr\ikl\frme\cdkk\aipt{--}\gs}.
The quantum mechanics
case has taken on renewed interest because superconformal
quantum mechanics may provide a dual description of
string theory on  $AdS_2$ \jm.

Most of the discussions so far have concerned relatively simple systems
either with small numbers of particles or exact integrability.
In this paper we consider a more general class of models with
$N$ particles.

We begin in section 2 with a bosonic sigma model with
an $N$-dimensional target space. It is shown that the model
has a nonlinearly-realized  conformal symmetry if and only if
the target space metric has a vector field $D^a\p_a$ whose Lie derivative obeys
\eqn\ldr{ {\cal L}_D g_{ab}=2g_{ab},}
and whose associated one-form is closed
\eqn\eok{d( D_adX^a)=0.}
Given \ldr\ and \eok\ it is shown that, in a Hamiltonian formalism, the
dilations are (roughly) generated by $D^aP_a$ while the special conformal
transformations are generated by
\eqn\rkh{K=\half D_aD^a .}
The conformal symmetry persists in the presence of a potential $V$ obeying
${\cal L}_D V =-2 V .$  A general class of examples is given.

In section 3 we turn to the supersymmetric case. The geometry
of Poincar\'{e}-supersymmetric quantum mechanics with a variety of
supermultiplets was discussed by Coles and Papadopoulos \cp.  We
restrict our attention to the case for which the multiplet
structure with respect to the Poincar\'{e} super-subgroup consists of
$N$ bosons $X^a$ with $N$ real superpartners $\lambda^a$. Such
multiplets arise in the reduction of two-dimensional chiral $(0,{\cal
N})$ multiplets, where $\cal N$ is the number of supersymmetries,
to one dimension, and give rise to what is sometimes referred to
as ``type $B$'' models (most of the literature concerns ``type $A$''
$({\cal N}/2,{\cal N}/2)$ multiplets). In section 3.1 we show that every
bosonic conformal model can be extended to an ${\cal N}=1B$ theory
with $Osp(1|2)$ superconformal symmetry provided the torsion obeys
certain constraints. In section 3.2 we consider ${\cal N}=2B$ and find that
the extension to
$SU(1,1|1)$ requires a complex structure $I$ with respect to which
$D$ must be holomorphic. $D^aI_a{^b}\p_b$ is found to generate a
$U(1)$ isometry. In section 3.3 we
first derive a simplified version of the conditions for ${\cal
N}=4B$ Poincar\'{e} supersymmetry with an $SU(2)$ $R$-symmetry as
first-order differential relations between the triplet of complex
structures $I^r$. We further show that an ${\cal N}=4B$ model has
a $D(2,1;\alpha)$ superconformal symmetry if the vector fields
$D^aI_a^{rb}\p_b$ generate an $SU(2)$ isometry group and obey
generalizations of the identities required for $SU(1,1|1)$. 
The parameter $\alpha$ is determined by the constant in the $SU(2)$ Lie
bracket algebra.
In
section 3.4 we construct a large class of ${\cal N}=4B$ theories
in terms of an unconstrained potential $L$. $D(2,1;\alpha)$ superconformal
symmetry then follows if $L$ is a homogeneous and $SU(2)$
rotationally-invariant function of the coordinates.  Related results in
four dimensions were recently discussed in~\dewit.

Throughout the
paper we use a Hamiltonian formalism. In appendix~\appq\ we give a
Lagrangian derivation of the supercharges used in the text. We
use real coordinates throughout the body of  the text, but
appendix B gives various useful formulae for the geometry and
supercharges in complex coordinates.  In appendix~\appwhkt\ we
discuss the conditions under which an ${\cal N}=4B$ geometry can be written
in terms of a potential $L$.

A primary motivation for this work is the expectation that
quantum mechanics on the five-dimensional
multi-black hole moduli space
is an ${\cal N}=4B$ theory with a $D(2,1;\alpha)$ superconformal symmetry
at low energies \jrmy .

\newsec{${\cal N}=0$ Conformally Invariant $N$-Particle Quantum Mechanics}
   In this section we find the conditions under which a general
$N$-particle quantum mechanics admits an $\slr$ symmetry.
We will adopt a Hamiltonian formalism,
and derive the conditions for the existence of appropriate
operators generating the symmetries.
The general Hamiltonian is%
\foot{The canonical momentum
$P_a=g_{ab}\dot X^b=-i\p_a$
obeys $[P_a,X^b]=-i{\delta_a}^b$ and $P_a^\dagger = {1 \over \sqrt{g}} P_a
\sqrt{g}$ (for the norm $(f_1,f_2) = \int d^N X
\sqrt{g} f_1^* f_2$).  In this and all subsequent expressions, the operator
ordering is as indicated.}
\eqn\mnhm{H=\half P_a^\dagger g^{ab} P_b +V(X).}
where $a,b=1,\ldots,N$.  We now determine the conditions under which the theory,
defined by equation~\mnhm, admits an $\slr$ symmetry.

We first look for a dilational symmetry of the general form
\eqn\yop{\eqalign{\delta_D X^a&=\e D^a (X),\cr
            \delta_D t&=2 \e t.\cr }}
This is generated by an operator
\eqn\hmt{{D}=\half D^a P_a + {\rm h.c.}}
which should obey
\eqn\rgls{[{D},H]=2iH.}
{}From the definitions \hmt\ and \mnhm\ one finds
\eqn\rgl{[{D},H]=-{i \over 2} P_a^\dagger ({\cal L}_D g^{ab}) P_b
               -{i } {\cal L}_D V
               -{i \over 4} \nabla^2 \nabla_a D^a,}
where ${\cal L}_D$ is the usual Lie derivative obeying: \eqn\oig{
{\cal L}_D
g_{ab}=D^cg_{ab,c}+{D^c}{_{,a}}g_{cb}+{D^c}{_{,b}}g_{ac}.}
Therefore, given a metric $g$ and potential $V$ a dilational
symmetry exists if and only if there exists a conformal killing
vector $D$ obeying \eqn\ldr{ {\cal L}_D g_{ab}=2g_{ab}} and
\eqn\rrp{{\cal L}_D V =-2 V .} Note that \ldr\ implies the
vanishing of the last term of equation~\rgl. A vector field $D$
obeying \ldr\ is known as a $homothetic$ vector field, and the
action of $D$ is known as a $homothety$.

Next we look for a special conformal symmetry
generated by an operator $K(X)$
obeying
\eqn\rgkl{[{D},K]=-2iK,}
and
\eqn\dhg{ [H,K]=-i{D}. }
Equations~\rgkl\ and~\dhg\ together with~\rgl\ is an $\slr$ algebra.
Equation~\rgkl\ is equivalent to
\eqn\kio{ {\cal L}_D K=2K,}
while \dhg\ can be written
\eqn\ius{D_adX^a =d K.}
Hence the one-form $D$ is exact.
One can
solve for $K$ as
\eqn\glo{K=\half g_{ab}D^aD^b.}
We shall adopt the phrase ``closed homothety'' to refer to
a homothety whose associated one-form is closed and exact.

An alternate basis of $\slr$ generators is
\eqn\rto{\eqalign{L_0&=\half(H+K),\cr  L_{\pm 1}&=\half(H-K\mp i {D}).}}
In this basis the generators obey the standard commutation relations
\eqn\rtl{\eqalign{[L_1,L_{-1}]&=2L_0,\cr [L_0,L_{\pm1}]&=\mp L_{\pm 1}.}}

The nature of these geometries can be illuminated by
choosing coordinates such that $(X^0)^2=2K$ and $g_{i0}=0$ for
$i=1,\ldots,N{-}1$. This is always locally possible away from the zeros of
$D$.  One then finds
\eqn\rux{\eqalign{ds^2&=(dX^0)^2+(X^0)^2 g_{ij}(X^k)dX^idX^j,\cr
 D^a\p_a&=X^0 {\partial \over \partial X^0}.}}
Hence, given $any$ metric $g_{ij}$ in $N-1$ dimensions, one can construct
a geometry with a closed homothety
in $N$ dimensions by dressing it with an extra radial
dimension.
Similar comments pertain to the potential $V$.

An alternate useful choice is dilational coordinates, in
which
\eqn\dga{D^a={2 \over
h}X^a,} where $h$ is an arbitrary constant.
These are related to the coordinates in
\rux\ by $X^{\prime 0}=(X^0)^{2/h},\;X^{\prime i}=(X^0)^{2/h} X^i$.
In such dilational coordinates one finds
\eqn\rtoq{X^a\p_ag_{bc}={h \over 2}{\cal
L}_Dg_{bc}-X^a{_{,c}}g_{ba}-X^a{_{,b}}g_{ac}=(h-2)g_{bc}.} Hence in
dilational gauge the metric components are homogeneous functions
of degree $h-2$. (It is not, however, the case that every homogeneous
metric admits an $\slr$ symmetry.) At this point $h$ can be changed by
transformations which take the coordinates to powers of themselves, and
so has no coordinate independent meaning. However, it turns out that for
${\cal N}=4B$ supersymmetry, a
preferred value of  $h$ is obtained in quaternionic coordinates, when such
coordinates exist and coincide with dilational gauge, as in the class of
examples considered in section 3.4.

In conclusion, the Hamiltonian \mnhm\ describes an $\slr$ invariant
quantum mechanics if and only if the metric admits
a closed homothety
\eqn\ldrf{\eqalign{ {\cal L}_D g_{ab}&=2g_{ab}, \cr
                       d (D_adX^a)&=0, }}
under which the potential transforms according to
\rrp.
\newsec{The Supersymmetric Case}

In the following we supersymmetrize the bosonic sigma model
by extending the boson $X^a$ to the supermultiplet $(X^a, \lambda^a)$
with $\lambda^a=\lambda^{a\dagger}$.
A number of other multiplets exist \cp\ which will not
be considered in the
following. Furthermore we will set the potential $V=0$.
An operator approach to a similar system can be found in
\jonghe.

\subsec{${\cal N}=1B$ Poincar\'{e} supersymmetry and $Osp(1|2)$
superconformal symmetry}
Let us supersymmetrize the bosonic sigma-model \mnhm\ for $V=0$  with
$N$ fermions $\lambda^\alpha $ where $\alpha=1,\ldots,N$ is a tangent space
index. These
obey the standard anticommutation relations%
\eqn\frm{\{\lambda^{ \alpha },\lambda^{\beta  }\}=\delta^{\alpha \beta  },}
and of course commute with $P_a$ and $X^b$.
It is convenient to make the field redefinitions
\eqn\frdef{\eqalign{\lambda^a&\equiv e^a{_\alpha }\lambda^\alpha,\cr
                     \Pi_a &\equiv
P_a-{i \over 2}\omega_{abc}\lambda^b\lambda^c+{i\over 2}c_{abc}\lambda^b
\lambda^c,}}
where $\omega$ is related to
the usual spin
connection by
$\omega_{abc} =\omega_a{^\beta  }{_\gamma}e_{b\beta}e_c^\gamma$.%
\foot{We note that $[\Pi_a,
\Pi_b]=-{\half}R^+_{abcd}\lambda^c\lambda^d$ where $R^+_{abcd}$ is constructed
from the connection $\Gamma^b_{ac}+c^b{_{ac}}$; $[P_a, \lambda^b]= -i
(\omega_a{^b}{_c} - \Gamma^{b}_{ac} )\lambda^c$ and $[\Pi_a,
\lambda^b]= i(\Gamma^{b}_{ac}+c^b{_{ac}}) \lambda^c$, where $\Gamma$ is the
Christoffel connection.  The Hilbert space can be viewed as  a
spinor (as is seen by identifying equation~\frm\ with the
$\gamma$-matrix algebra) and $\Pi_a$ as the covariant derivative (with
torsion $c$)
on Hilbert space states.}\

A supercharge can then be constructed as%
\foot{Despite the
non-hermiticity of $\Pi_a$, this expression is
hermitian with the indicated operator ordering.}
\eqn\scq{Q=\lambda^a\Pi_a -{i \over 3} c_{abc} \lambda^a \lambda^b \lambda^c,}
where $c$ is a 3-form, which at this point is arbitrary.  A derivation of
the supercharge from a supersymmetric Lagrangian is given in
appendix~\appq.  The supercharge obeys
\eqn\qs{ \{Q,Q\}= 2 H ,}
where the bosonic
part of $H$ agrees with \mnhm\ for $V=0$.

We wish to extend this ${\cal N}=1B$ Poincar\'{e}-superalgebra to the
$Osp(1|2)$ superconformal algebra whose non-vanishing commutation relations are
\eqn\ccr{\matrix{\hfill [H,K]\smaleq -iD, \hfill &&&
                 \hfill [H,D]\smaleq -2iH, \hfill &&&
                 \hfill [K,D]\smaleq 2iK,\hfill \linesp
\hfill \{Q,Q\}\smaleq 2H, \hfill &&&
   \hfill [Q,D]\smaleq -iQ, \hfill &&&
   \hfill [Q,K]\smaleq -iS, \hfill \linesp
\hfill \{S,S\}\smaleq 2K, \hfill &&&
   \hfill [S,D]\smaleq iS, \hfill &&&
   \hfill [S,H]\smaleq iQ, \hfill \linesp
\hfill \{S,Q\}\smaleq D.\hfill }}
As before, the bosonic subalgebra requires a closed homothety.
The new supercharge can then be constructed
as
\eqn\rth{S=i[Q,K]=\lambda^aD_a.}
with $K$ given by \glo.
The $\anti{S}{Q}$ anticommutator is then used to find
\eqn\qsa{\anti{S}{Q} = D = \half ( D^a \Pi_a + {\rm h.c.} ).}
Then, $\com{S}{D}=iS$ is satisfied, but
the $[Q,D]$ commutator is
\eqn\dqc{[Q,D]=-iQ-ic_{abc}D^a \lambda^b P^c +\ol{3}.}
Agreement with \ccr\ then requires $c$ to be orthogonal to $D$:
\eqn\cls{D^ac_{abc}=0.}
Given \cls, the full commutator becomes
\eqn\dqsc{[Q,D]=-iQ-{1\over6} \lambda^a\lambda^b\lambda^c({\cal L}_D-2) c_{abc}.}
We therefore demand that $c$  transform under dilations as
\eqn\psst{{\cal L}_Dc_{abc}=2c_{abc}.}
The remaining commutators in \ccr\ then follow from the Jacobi identities,
with no further constraints
on the geometry.

In summary any ${\cal N}=0$ conformal quantum mechanics can be
promoted to $Osp(1|2)$, but the torsion $c$ appearing in the
supercharges must obey
\eqn\rtok{\eqalign{D^ac_{abc}&=0,\cr {\cal L}_Dc_{abc}&=2c_{abc}.} }

\subsec{${\cal N}=2B$ Poincar\'{e} supersymmetry and $SU(1,1|1)$ superconformal
symmetry}

${\cal N}=2B$ supersymmetry requires a complex
structure $I$ and a
hermitian metric on the target space \cp. The relevant formulae are
simplest in complex coordinates. However complex
coordinates are less useful in the extension to the $4B$ case (which has an
$SU(2)$ triplet of complex structures) considered in the next subsection.
Accordingly we continue with real coordinates, but give the complex
version in  appendix B.

The second supercharge is given by
\eqn\sgn{\tilde Q=\lambda^aI_a{^b}\Pi_b
-{i\over2} \lambda^a I_a{^b} c_{bcd} \lambda^c \lambda^d
-{i\over6} \lambda^a \lambda^b \lambda^c I_a{^d} I_b{^e} I_c{^f} c_{def}
-{i\over2} \lambda^a c_{abc} I^{bc}.}
A derivation is given in appendix B.
Whereas $c$ is unconstrained for ${\cal N}=1B$, for ${\cal N}=2B$
the vanishing of $\{ \tilde Q,Q \}$ requires~\cp
\eqn\csc{\nabla^+_{(b}I_{c)}{^a}=0,}
where the torsion connection $\nabla^+$ involves
the Christoffel connection plus the torsion $c$ as $\Gamma^b_{ac}+c^b{_{ac}}$
In complex coordinates \csc\ can be solved for the
$(1,2)$ part of $c$ as
\eqn\rfo{c|_{(1,2)}=-{i \over 2}\bar \p J,}
with
\eqn\jnku{J=\half I_a{^c}g_{bc}dX^a\wedge dX^b.}
The $(0,3)$ part of $c$ must be closed under $\bar{\p}$ but is otherwise
unconstrained, and the $(2,1)$ and $(3,0)$
parts are obtained by complex conjugation.

We wish to promote the ${\cal N}=2B$ algebra to $SU(1,1|1)$. This
involves an additional bosonic generator $R$ which is the generator of the
$R$ symmetry group of the ${\cal N}=2B$ subalgebra.
The non-vanishing commutation relations are given by \ccr, an identical
set of relations with both $Q$ and $S$ replaced by $\tilde Q$ and $\tilde
S$, together with
\eqn\ccrb{\matrix{
\hfill \{\tilde Q,S \}\smaleq R, \hfill &&&
   \hfill \{\tilde S,Q \}\smaleq -R,\hfill \linesp
\hfill [R,Q]\smaleq -i\tilde Q,\hfill &&&
   \hfill [R,\tilde Q]\smaleq i Q,\hfill \linesp
\hfill [R,S]\smaleq -i\tilde S,\hfill &&&
   \hfill [R,\tilde S]\smaleq iS. \hfill }}

As before closure of the algebra
requires that the geometry must admit a closed homothety,
as well as the constraints \rtok\ on $c$.
Commutation of the supercharges with $K$ leads to the
superconformal charges
\eqn\ydop{\eqalign{S&=\lambda^aD_a,\cr
          \tilde S &=\lambda^aI_a{^b}D_b.}}

Obtaining the correct commutator $[D,\tilde Q]=i\tilde Q $  requires that
the action of $D$ preserves the complex structure:
\eqn\sdt{{\cal L}_{D}I_a{^b}=0.}
This is equivalent to the statement that $D$ acts holomorphically.
Alternate forms of \sdt\ are
\eqn\ips{\eqalign{
{D}^f I_f{^a}
c_{abc} I_d{^b} I_e{^c} &= {D}^f I_f{^a} c_{abc}; \cr
\p_i D^{\bar{j}} &= 0 .}}
\def\ugly{\ips}

It follows from \ips, together with \cls\ and \rfo\ that
$\tilde{D}^a=D^b I_b{^a}$ generates a holomorphic isometry
\eqn\trps{\eqalign{{\cal L}_{\tilde{D}}I_a{^b}&=0,\cr {\cal
L}_{\tilde D}g_{ab}&=0,}} as expected from $\com{R}{H}=0$.
Moreover the $(2,1)$ part of the torsion $c$ is annihilated by
${\cal L}_{\tilde D}$ while the $(3,0)$ part has weight $-2i$.

$R$ is determined from the commutator of $Q$ and $\tilde S$ as
\eqn\yil{R=\tilde{D}^a\Pi_a-i I_{ab} \lambda^a\lambda^b -
i \tilde{D}^a c_{abc} \lambda^b \lambda^c \,} where we
used equation~\ugly. One finds
\eqn\ldc{\eqalign{[R,\lambda^a]&=-i(I^a{_b}
  +\tilde{D}^a{_{,b}})\lambda^b,\cr
[R,D^a]&=-i\tilde D^b D^a{_{,b}} .}} In
complex coordinates and
dilational gauge
$hD^a=2X^a$, when such coordinates exist, this reduces to
\eqn\ldc{\eqalign{[R,\lambda^a]&=-i(1+{2 \over h})
\lambda^bI_{b}{^a},\cr [R,X^a]&=-{2i\over h}X^bI_{b}{^a}.}}
Notice that $R$ commutes with $\lambda$
in complex coordinates with
$h=-2$.

 All the remaining commutators~\ccrb\
and~\ccr\ are satisfied without any additional constraints.

In summary,  there is an $SU(1,1|1)$ symmetry if and only if,
in addition to the $Osp(1|2)$ constraints
\ldrf\ and \rtok, and the ${\cal N}=2B$ constraints, $D$ preserves the
complex structure:
\eqn\trp{\eqalign{{\cal L}_{{D}}I_a{^b}&=0. \cr }}
It further follows that $\tilde{D}^a=D^b
I_b{^a}$ generates a holomorphic isometry.

\subsec{${\cal N}=4B$ Poincar\'{e} supersymmetry and $D(2,1;\alpha)$
superconformal symmetry}

\subsubsec{Remarks on ${\cal N}=4B$ Poincar\'{e} supersymmetry}

Extending the algebra to include 4 supersymmetries requires 3
complex structures
$I^r$, $r=1,2,3$.
With each $I^r$ one can associate a generalized exterior derivative%
\eqn\dfer{d^r=dX^a I^r_a{^b} \nabla^r_b\wedge ,}
where the connection $\Omega^r$ appearing in $\nabla^r$ is%
\foot{$\Omega^r$ defined in this way gives a connection acting on
forms as described but not on general tensors.}
\eqn\rot{{\Omega^{r a}}_{bc}=-I^{ra}_d\p_cI^{rd}_b.}
One of the conditions for ${\cal N}=4$ supersymmetry found
in  \refs{\cp,\gps} can be expressed
\eqn\rxm{\{ d^r, d^s \} =0.}
These are the vanishings of the Nijenhuis tensors
and concomitants.%
\foot{So, Theorem 3.9 of~\intya\ implies that the vanishing of any two
of these equations
yields the vanishing of all six.}
In complex coordinates adapted to $I^r$,
$\Omega^r$ vanishes and $d^r{=}i(\p{-}\bar{\p})$.
Equation~\rxm\ further implies
\eqn\rxd{\{d^r,d\} = 0.}
Additional requirements for
supersymmetry discussed in \refs{\cp,\gps} are
\eqn\dtp{g_{ab}=I^{rc}_aI^{rd}_b g_{cd}\ (\forall\ r),}
\eqn\ioi{ \{I^r,I^s \}=-2\delta^{rs},}
\eqn\cvb{\p_{[a}(I_b^{re}c_{|e|cd]})-2I_{[a}^{re}\p_{[e}c_{bcd]]}=0,}
\eqn\cvc{\nabla^+_{(b}I^{ra}_{c)}=0.}
In this last equation, we used the covariant derivative with torsion
$\nabla^+$ defined
just below equation~\csc.

The commutators of $I^r$ are related to the $R$-symmetry group.
We shall consider the $SU(2)$ case\foot{We have employed an obvious
summation convention in this equation. We hope that it will be clear from
the context when repeated indices should or should not
be summed over.}
\eqn\juh{[I^r,I^s ]=2\epsilon^{rst}I^t.}
This case is sometimes referred to as ${\cal N}=4B$ supersymmetry,
and arises in the reduction of $(0,4)$ supersymmetry from two dimensions.

We now show, defining the two-forms
\eqn\jnk{J^r=\half I^r_a{^c}g_{bc}dX^a\wedge dX^b,}
that the necessary and sufficient
conditions for ${\cal N}=4B$
supersymmetry can be recast in the simpler form
\eqn\rxm{\{ d^r, d^s \} =0,}
\eqn\dtp{g_{ab}=I^{rc}_aI^{rd}_b g_{cd}\ (\forall\ r),}
\eqn\juh{I^rI^s =-\delta^{rs}+\epsilon^{rst}I^t,}
\eqn\typ{d^1J^1=d^2J^2=d^3J^3.}
Note that the last two conditions \cvb\ and \cvc\
which involve the torsion $c$ have been replaced by the
condition \typ\ which is independent of $c$.
Let us write the torsion appearing in
\cvc\ as
\eqn\cdf{c= \half d^3J^3 +e}
for some three-form $e$.
It can be checked that the
torsion connection with $e$ set to zero is the unique such connection
annihilating
$I^3$, and therefore has holonomy contained in $U(N/2)$. 
It follows that, in complex coordinates adapted to $I^3$,
the condition \cvc\ for $r=3$ reduces to
\eqn\chft{e_{i \bar j \bar k}=e_{\bar i  jk}=0.}
(This is the argument that led to equation~\rfo.)
On the other hand, adding the $r=1$ plus or minus $i$ times the
$r=2$ component of \cvc\ yields
\eqn\cft{e_{ijk}=e_{\bar i \bar j \bar k}=0.}
We conclude that $e=0$ and $c= \half d^3J^3$. By symmetry we must also have
$c= \half d^1J^1$ and $c= \half d^2J^2$, from which \typ\ follows. Conversely
given \typ, adding the torsion $c= \half d^3J^3$ to the Christoffel connection
implies \cvc.  It can be further checked that this choice of $c$ satisfies
\cvb.

This single choice of
torsion connection annihilates all three complex structures
\eqn\cvyc{\nabla^+_{b}I^{r}_{c}{^a}=0.}
In fact the condition \cvyc\ is equivalent to \typ.
It differs from \cvc\ by the absence of symmetrization
but is nevertheless equivalent for ${\cal N}=4B$.
Equation~\cvyc\ is referred to in \gps\ as the
weak HKT (hyperk\"{a}hler with torsion) condition.
We have shown that ${\cal N}=4B$ (which includes the condition \juh )
implies weak HKT.

\subsubsec{Extension to $D(2,1;\alpha)$ superconformal symmetry}

We now turn to superconformal symmetry. It turns out that the
relevant supergroup is $D(2,1;\alpha)$, where the parameter
$\alpha{\neq}{-}1$ will be  determined by the geometry.
In order to write down the commutators, it is convenient to
define the four-component supercharges $Q^m = (Q^r,Q)$ and
$S^m=(S^r,S)$ for $m=1,2,3,4$; these transform in the $(2,2)$ of
the $SU(2)\times SU(2)$ $R$-symmetry group of ${\cal N}=4B$.
Operators $Q^m$, $S^m$, $H$, $D$, $K$ and $R_\pm^r$ (to be
described) then comprise the $D(2,1;\alpha )$ algebra.  The non-vanishing
commutators are
$$\matrix{\hfill [H,K]\smaleq-iD,\hfill &&&
   \hfill [H,D]\smaleq-2iH,\hfill &&&
   \hfill [K,D]\smaleq2iK,\hfill \linesp
\hfill \{{ Q}^m,{ Q}^n\}\smaleq2H\delta^{mn},\hfill &&&
\hfill [{ Q}^m,D]\smaleq-i{ Q^m},\hfill &&&\hfill [{ Q^m},K]\smaleq
   -i{ S^m},\hfill \linesp
\hfill \{{ S^m} ,{ S}^n\}\smaleq\hfill 2K\delta^{mn},&&&
   \hfill [{ S^m},D]\smaleq i{ S^m},\hfill &&&
   \hfill [{ S^m},H]\smaleq i{ Q^m},\hfill \linesp
\hfill [R_\pm^r,{ Q^m}]\smaleq it^{\pm r}_{mn}Q^n,\hfill &&&
  \hfill [R_\pm^r,{ S^m}]\smaleq it^{\pm r}_{mn}S^n,\hfill &&&
  \hfill [R_\pm^r,R_\pm^s]\smaleq i\epsilon^{rst}R_\pm^t \hfill}$$
\eqn\ccrd{\{{ S^m,Q^n}\}=D\delta^{mn}-{4 \alpha \over 1+\alpha }
  t^{+r}_{mn}R_+^r-
{4 \over 1+\alpha }t^{-r}_{mn}R_-^r.}
The $t^\pm$ matrices
defined by \eqn\tpm{t^{\pm r}_{mn}\equiv
\mp\delta^r_{[m}\delta^4_{n]}+\half\epsilon_{rmn}} obey
\eqn\tpmo{[t^{+r},t^{-s}]=0, \qquad [t^{\pm r},t^{\pm
s}]=-\epsilon^{rst}t^{\pm t}, \qquad \{ t^{\pm r},t^{\pm s} \}
=-\half \delta^{rs}.} Notice that when $\alpha=0$ or
$\alpha=\infty$, one of the two $SU(2)$s can be decoupled, and
there is an $SU(1,1|2)$ subalgebra.

Since $D(2,1|\alpha )$ has three $SU(1,1|1)$ and one ${\cal N}=4B$
subalgebra, the (previously discussed) conditions on the geometry
for the existence of those subgroups can all be assumed. In
particular, $D$ must now be holomorphic with respect to all three
complex structures \eqn\fgt{{\cal L}_{D}I^{rb}_a=0.} Expressions
for $Q^r$ and $S^r$ are then of the $SU(1,1|1)$ forms \sgn\ and
\ydop\ with $I$ replaced by $I^r$.  Somewhat lengthy expressions for
$R^r_\pm$ as a
function of $\alpha$ then follow from linear combinations of
$\{Q^m,S^n\}$ anticommutators as determined by \ccrd.%
\foot{In principle, we should treat $\alpha=0$ or $\infty$ as special
cases.  In fact, $\alpha=\infty$ cannot be realized with the supermultiplet
we are considering.
For $\alpha=0$, the logic is slightly different but
the results are the same.}
Obtaining
properly normalized $SU(2)$ algebras for the operators $R^r_\pm$
so determined requires 
\eqn\ret{ [{\cal L}_{D^r},{\cal
L}_{D^s}]={4 \over h}\epsilon^{rst}{\cal L}_{D^t},} with
$D^r=D^aI^{rb}_a\p_b$ and \eqn\hap{h=-2\alpha-2.}
Equation \ret\ can be
taken as the definition of the constant $h$.%
\foot{Note that the two excluded values $\alpha=-1$ and
$\alpha=\infty$, correspond respectively to $h=0$ and $h=\infty$, for which
the algebra~\ret\ is clearly singular.} 
Since the
normalization of $D^r$ is fixed in terms of $D$, $h$ is a
coordinate-invariant parameter associated to the geometry.

Reproducing the proper $[R,Q]$ commutators leads to the stronger
requirement
\eqn\str{{\cal L}_{D^r}I_a^{sb}={4 \over
h}\epsilon^{rst}I_a^{tb}.} In fact \str\ (including $r=s$) implies
both \ret\ and \fgt. Using \str\ one then finds $R^r_\pm$ are
given by \eqn\yh{R^r_-=-{h\over4} D^{ra}\Pi_a + i {h-2 \over 8} \lambda^a
I^r_a{^b} \lambda_b + i {h\over4} \lambda^a \lambda^b D^{rc} c_{cab}}
 \eqn\ysh{R^r_+={i \over
4}\lambda^aI^r_a{^b}\lambda_b.}
The torsion $c$ can be eliminated from \yh\ using the identity $D^{rc}
c_{cab} = \half(d^rdK)_{ab} - J^r_{ab}$.

Using the Jacobi identity, the remaining commutators follow with
no further constraints on the geometry.

We note that equations~\str\ and~\cvyc\ imply
\eqn\tyi{\eqalign{2(h+2)J^r&={h}(d^rdK-\half\epsilon^{rst}d^sd^t K),\cr
4(h+2)c&=-{h}d^1 d^2 d^3 K.}}
Properties of $d^r$ and $d$ then imply equation \typ, which thus needs not be
taken as a further condition.

We also find, in quaternionic coordinates and dilational gauge, when such
coordinates exist, that
\eqn\frg{\matrix{\cr
\hfill [R^r_-,\lambda^a]\smaleq 0, \hfill &&&
\hfill [R^r_-,X^a]\smaleq{i \over 2} X^bI^{ra}_b,\hfill \linesp 
\hfill [R^r_+,\lambda^a]\smaleq{i\over 2} \lambda^bI^{ra}_b,\hfill &&&
\hfill [R^r_+,X^a]\smaleq0.\hfill
}}

In summary, a quantum mechanical theory has ${\cal N}=4B$
supersymmetry if and only if the complex structure and metric obey
equations~\rxm--\typ. The torsion $c$ is then uniquely determined
as \eqn\ceq{c=\half d^3J^3.} A $D(2,1;\alpha)$ symmetry arises if
and only if in addition there is a vector field $D$ obeying
\eqn\ldf{\eqalign{ {\cal L}_D g_{ab}&=2g_{ab}, \cr
              d (D_adX^a)&=0,\cr
              {\cal L}_{D^r}I^{s}_a{^b}&={4 \over
              h}\epsilon^{rst}I^t_a{^b},\cr
              {\cal L}_{D^r}g_{ab} &= 0,\cr
 }}
where $D^{rb}=D^aI^{rb}_a$ and $h$ is a constant characterizing
the geometry. The parameter $\alpha$ in the superconformal algebra
is related to the constant ${ h}$ in \ldf\ by
\eqn\arl{\alpha =
-{h+2 \over 2}.}

\subsec{Examples of $D(2,1;\alpha)$ Quantum Mechanics}

In this subsection, we show that a large class of examples of quantum
mechanical systems with
$D(2,1;\alpha)$ symmetry (and an integrable quaternionic
structure) can be constructed from a potential $L$. In an ${\cal
N}=2$ superspace formalism (not described here, but similar to the
ones in~\refs{\ghr,\dps}) $L$ turns out to be the superspace
integrand.

$\IR^4$ has an obvious $SU(2)$ triplet of complex structures
associated to self-dual two-forms obeying \juh. Let $I^r$ be the
generalizations to $\IR^{4N}$. We may then define a triplet of
fundamental two-forms by \eqn\tpr{J^r={1 \over 8}(2d^r d L
-\epsilon^{rst}d^sd^t L).} It follows immediately from this
definition and $\{d^r,d^s\}=0$ that the $J^r$ obey \typ. Moreover
the associated metric $g_{ab}=I^r_b{^c}J^r_{ac}\;(\forall\ r)$
can be written (in a coordinate system in which the $I^r$ are
constant) \eqn\gol{g_{ab} = {1 \over 4}\bigl(\delta_a^c \delta_b^d
+ I^r_a{^c} I^r_b{^d} \bigr) \p_c \p_d L.} This expression is
manifestly hermitian. In other words for any $L$ we can construct
an ${\cal N}=4B$ quantum mechanics.\foot{Although one may wish in
addition to impose positivity of the metric $g$, which further
constrains $L$.} It is natural to ask whether or not every weak
HKT geometry is described by some potential $L$. This is related
to the integrability of the quaternionic structure, as discussed
in appendix~\appwhkt.

The full $D(2,1;\alpha)$ symmetry follows by imposing
\eqn\rtos{X^a\p_aL=h L,} where $h$ is an arbitrary constant and
\eqn\rtoi{X^a I^{rb}_a \p_b L=0.} The first condition implies that
$L$ is a homogeneous function of degree $h$ on $\IR^{4N}$, while
the second states that it is invariant under $SU(2)$ $R$-symmetry
rotations. These conditions manifestly ensure the existence of the
required homothety
\eqn\pol{D^a\p_a={2 \over h} X^a\p_a }
as well as the
$SU(2)$ isometries. Remarkably, it follows from \rtos\ and \rtoi\
with a little algebra that $D$ is automatically a {\it closed}
homothety, \eqn\cld{D_a dX^a = {(h+2) \over2 h} (\p_a L)\, dX^a .}
As discussed in section 2 this implies the existence of special
conformal transformations generated by \eqn\mtnno{K=\half
g_{ab}D^a D^b= {(h +2)\over 2h} L.}  In fact, all the requirements of~\ldf\ 
are automatically satisfied with these conditions, and so indeed the full
$D(2,1;\alpha=-{h+2\over2})$
algebra is obtained.

The conditions \rtos\ and \rtoi\ are sufficient but not necessary
to insure $D(2,1;\alpha)$ invariance. More generally one could add
to the right hand side anything which is in the kernel of the
second-order differential operator in \tpr. This is especially
relevant for the interesting case $h=-2$, for which
equations~\cld\ and~\mtnno\ show that the metric is otherwise
degenerate. 
An example of this will appear in \jrmy.

The simplest case is \eqn\scse{L=\half \delta_{ab}X^aX^b,} where
$a,b=1,\ldots,4N$. This has $h=2$. The metric is then simply the flat
metric on $\IR^{4N}$
\eqn\dfe{ds^2=\delta_{ab}dX^adX^b,}
while
the torsion $c$ vanishes. The generators of $D(2,1;-2)\sim Osp(4|2)$ are then
\eqn\drt{\matrix{\hfill H\smaleq\half P^aP_a,\hfill&&&
\hfill K \smaleq \half X^aX_a, \hfill&&&
\hfill D \smaleq X^aP_a,\hfill\linesp
\hfill Q \smaleq \lambda^aP_a,\hfill&&&
\hfill Q^r \smaleq \lambda^aI^{rb}_aP_b,\hfill\linesp
\hfill S \smaleq \lambda^aX_a,\hfill&&&
\hfill S^r \smaleq \lambda^aI^{rb}_aX_b,\hfill\linesp
\hfill R_-^r \smaleq -{1 \over 2}X^aI^{rb}_aP_b,\hfill&&&
\hfill R_+^r \smaleq {i\over 4} \lambda^aI^{rb}_a\lambda_b.\hfill}}

\centerline{\bf Acknowledgements}

  We have benefitted from useful conversations with
R.\ Britto-Pacumio,
J.\ Gutowski, J.\ Maldacena, 
A.\ Maloney, M.\ Spalinski,
M.\ Spradlin, P.\ Townsend,
A.\ Volovich
and especially G.\ Papadopoulos.
This work was supported in
part by an NSERC PGS B Scholarship
and DOE grant DE-FGO2-91ER40654.

\appendix{\appq}{Lagrangian Derivation of the Supercharges}

In this section we derive the supercharges used in the body of the text,
from the component action~\refs{\cp,\gps}
\eqn\lags{S = \int dt \left\{ {1\over2} g_{ab} \dot{X}^a \dot{X}^b
+ {i\over2} \lambda^a \left(g_{ab}{D\lambda^b\over dt} - \dot{X}^c
c_{abc} \lambda^b \right)
- {1\over6} \p_d c_{abc} \lambda^d \lambda^a \lambda^b \lambda^c,\right\}}
where the covariant derivative is
\eqn\lag{{D\lambda^b\over dt} \equiv \dot{\lambda}^b + \dot{X}^c
\Gamma^b_{cd} \lambda^d,}
with $\Gamma$ the Christoffel connection,
and we use dots to denote time derivatives.  Although we have, for ease of
manipulation, written the
fermions with spacetime indices, in deriving commutators it is better to
use $\lambda^\alpha$, where $\alpha$ is a tangent index, because,
unlike $\lambda^a$,
it will commute with the momentum conjugate to $X$.
In terms of $\lambda^\alpha$, the kinetic term for
the fermions is
\eqn\kin{{i\over2} g_{ab} \lambda^a {D\lambda^b\over dt}
= {i\over2} (\delta_{\alpha \beta} \lambda^\alpha \dot{\lambda}^\beta +
\dot{X}^c \omega_{c \alpha \beta  } \lambda^\alpha  \lambda^\beta)  ,}
and the momentum conjugate to $X$ is
\eqn\mom{P_a = g_{ab} \dot{X}^b + {i\over2}(\omega_{abc}-c_{abc}) \lambda^b
\lambda^c,}
or, using the definition in~\frdef,
\eqn\momb{\Pi_a = g_{ab} \dot{X}^b.}

The action~\lag\ is invariant under
the supersymmetry transformation
\eqn\susya{\delta_\epsilon X^a = -i \epsilon \lambda^a \qquad \qquad
\delta_\epsilon \lambda^a = \epsilon \dot{X}^a,}
where $\epsilon$ is a real anticommuting parameter.
Note that
\eqn\chka{\com{\delta_\epsilon}{\delta_\eta} = -2 i \eta \epsilon {d \over
dt},}
as required of a supersymmetry transformation.
It is straightforward
to compute the \nother\ charge corresponding to this symmetry; we
find%
\eqn\sca{Q = \lambda^a \Pi_a - {i \over 3} c_{abc} \lambda^a \lambda^b
\lambda^c,}
which is the origin of equation~\scq.  Actually,
the \nother\ procedure determines the charge
only up to operator ordering. We have fixed this ambiguity by demanding
hermiticity and target space covariance.

\appendix{B}{${\cal N}=2B$ Supersymmetry in Complex Coordinates}

In this appendix we revisit the ${\cal N}=2B$ supersymmetry of
section 3.2 in complex coordinates, which simplifies the formulae
and calculations. Equation \rfo\ for the $(1,2)$ part of the
torsion is \eqn\tid{c_{i \bar j \bar{k}} =
 g_{i[\bar j,\bar k]}.}  The $(3,0)$
part of the torsion is constrained by the relation
\eqn\rfv{c_{[l},_{ijk]}=0.} Identities required of $D^a$ are
\eqn\dfb{\eqalign{D^kg_{i \bar j,k}+D^k{_{,i}}g_{k \bar j}&=g_{i
\bar j},\cr D^ic_{ijk}&=0,\cr D^{\bar i} c_{\bar i jk}&=0,\cr
D^{\bar i} c_{\bar i \bar jk}&=-D^{ i} c_{ i \bar jk}.}}

It is convenient to define a complex supercharge
\eqn\scbb{{\cal Q} = \half(Q - i \tilde{Q}) \qquad \qquad \bar{\cal Q} =
\half(Q+i\tilde{Q}).} 
$\cal Q$ can be determined by the
requirement $Q={\cal Q}+\bar{\cal Q}$ together with
\eqn\dty{\{{\cal Q}, \lambda^{k}\}=0.} 
Equation~\dty\ is a
manifestation of the separation of the holomorphic and antiholomorphic
parts of the theory. One finds 
\eqn\sccc{{\cal Q} = \lambda^{{i}} \Pi_{{i}}
- i c_{ij\bar{k}} \lambda^i\lambda^j\lambda^{\bar{k}} 
- {i\over3} c_{ijk} \lambda^{i}\lambda^{j} \lambda^{k} 
- i c^{j}{_{jk}}\lambda^{k}.} 
Using 
\eqn\pda{(\lambda^{k}\Pi_{k})^\dagger = \lambda^{\bar{k}} \Pi_{\bar{k}},} 
one finds the hermitian conjugate is 
\eqn\sccc{\bar{\cal Q} = \lambda^{\bar{i}} \Pi_{\bar{i}} 
- i c_{\bar{i}\bar{j}k} \lambda^{\bar{i}}\lambda^{\bar{j}} \lambda^k
- {i\over3} c_{\bar{i}\bar{j}\bar{k}} \lambda^{\bar{i}}\lambda^{\bar{j}}
\lambda^{\bar{k}} - {i} c^{\bar{j}}{_{\bar{j}\bar{k}}}\lambda^{\bar{k}}.} 
After reordering the operators these expressions
agree with that for $\tilde Q$ in the text.

It is straightforward, though quite tedious, to obtain this
expression for $\tilde{Q}$ as
the (hermitian) \nother\ charge for the second supersymmetry
\eqn\susyb{\tilde{\delta}_\epsilon X^a = -i \epsilon I_b{^a} \lambda^b
\qquad \qquad
\tilde{\delta}_\epsilon \lambda^a = -\epsilon [I_b{^a} \dot{X}^b - i
\lambda^c (\p_c I_b{^a})\lambda^b].}
We note that the extra term in the transformation of $\lambda$ not only
appears naturally from the ${\cal N}=1$ superspace formulation of~\cp, but
also is necessary to obtain the algebra
\eqn\chkb{\com{\tilde{\delta}_\epsilon}{\delta_\eta} = 0 \qquad \qquad
\com{\tilde{\delta}_\epsilon}{\tilde{\delta}_\eta} = -2i \eta \epsilon
{d\over dt},}
for $I$ a complex structure with vanishing Nijenhuis tensor.  Note that
equation~\susyb\ implies equation~\dty.

\appendix{\appwhkt}{More on the Geometry of Weak HKT Manifolds}
    In this appendix we will show that, given a weak HKT manifold with
integrable complex structures,
we can find a potential $L$.  We will prove this shortly but first
we should elaborate on the assumption that the
quaternionic structure is integrable.

It is well known that, for almost complex
manifolds, the almost complex structure is integrable---that is, there exists a
coordinate system in which the components of the almost complex structure are
constant---if and only if the Nijenhuis tensor vanishes.  The analogous
statement is {\it not} true for quaternionic manifolds; rather, the
vanishing of the six Nijenhuis concomitants on an almost quaternionic
manifold only
guarantees the integrability of any
one complex structure.  To see this (see also \hp), suppose that we work in
a complex
coordinate system adapted to $I^3$.  Then, $I^1$ and $I^2$ have only mixed
indices ({\it i.e.}, as forms they are $(2,0)\oplus(0,2)$ forms).  Now consider the
connection~\refs{\obata,\intya}
\foot{This is not identical to equation~\rot.
Equation~\rot\ was
written in a general coordinate system, whereas the following equation is
written in coordinates adapted to the $I^3$.  Thus, $C$ is a connection,
provided one restricts oneself to holomorphic coordinate transformations,
for $C$ depends implicitly on $I^3$, while $\Omega^1$ did not.}
\eqn\purecon{C^k_{ij} = I^1_{\bar{l}}{^k} \partial_i I^1_{j}{^{\bar{l}}}
\qquad {\rm and} \qquad {\rm c.c.}}
The vanishing of the Nijenhuis tensor implies that $C^k_{ij}$ is actually a
symmetric connection.
Furthermore, this connection vanishes in a basis (if one exists) in which
$I^1$ and $I^3$ are simultaneously constant,
and so its curvature
tensor vanishes in such a basis.  Thus, a necessary condition for
integrability of the
quaternionic structure is the vanishing of the curvature associated with
the connection~\purecon.  Obata~\obata\ has shown that this is also a
sufficient condition.

If we assume integrability of the quaternionic structure, then we can,
without loss of generality, work in a basis in which the complex structures
are given by
\eqn\intcplxstr{\eqalign{
I^1 &=
    i d\bar{w}^A \otimes {\partial \over \partial z^A}
  - i d\bar{z}^A \otimes {\partial \over \partial w^A}
  - i dw^A \otimes {\partial \over \partial \bar{z}^A}
  + i dz^A \otimes {\partial \over \partial \bar{w}^A} \cr
I^2 &=
      d\bar{w}^A \otimes {\partial \over \partial z^A}
  -   d\bar{z}^A \otimes {\partial \over \partial w^A}
  +   dw^A \otimes {\partial \over \partial \bar{z}^A}
  -   dz^A \otimes {\partial \over \partial \bar{w}^A} \cr
I^3 &=
    i dz^A \otimes {\partial \over \partial z^A}
  + i dw^A \otimes {\partial \over \partial w^A}
  - i d\bar{z}^A \otimes {\partial \over \partial \bar{z}^A}
  - i d\bar{w}^A \otimes {\partial \over \partial \bar{w}^A},
}}
where we have split up the complex coordinates into
two sets $(z^A,w^A)$, $A=1,\ldots,{N\over4}$.
Hermiticity of the metric with respect
to $I^1$ (we do not get any additional information from $I^2$) implies that
\eqn\herm{g_{z^A \bar{z}^B} = g_{w^B \bar{w}^A}; \qquad \qquad
          g_{z^A \bar{w}^B} = -g_{z^B \bar{w}^A} = g_{z^{[A} \bar{w}^{B]}}.}
The condition that $d^1 J^1 = d^3 J^3$ then becomes%
\foot{Again, we do not get any additional information from $J^2$, since
$d^3 J^3$ is $(1,2)\oplus(2,1)$ and the
(2,1) and (1,2) parts of $d^2 J^2$ are trivially equal to those of $d^1
J^1$ and the (0,3) and (3,0) parts of $d^2 J^2$ are just minus those of
$d^1 J^1$.}\
\eqn\whktifint{\eqalign{g_{z^{[A} \bar{w}^B,\bar{w}^{C]}} = 0 \qquad & \qquad
                        g_{w^{[A} \bar{z}^B,w^{C]}} = 0 \cr
                        g_{w^{[A} \bar{z}^B,\bar{z}^{C]}} = 0 \qquad & \qquad
                        g_{z^{[A} \bar{w}^B,z^{C]}} = 0 \cr
g_{z^{[A} \bar{z}^{|C|},\bar{w}^{B]}} - \half g_{z^A \bar{w}^B,\bar{z}^C} = 0
\qquad & \qquad
g_{z^{A} \bar{z}^{[B},w^{C]}} + \half g_{w^B \bar{z}^C,z^A} = 0 \cr
g_{z^{A} \bar{z}^{[B},\bar{z}^{C]}} - \half g_{w^B \bar{z}^C,\bar{w}^A} = 0
\qquad & \qquad
g_{z^{[A} \bar{z}^{|C|},z^{B]}} + \half g_{z^A \bar{w}^B,w^C} = 0.}}

The first and second lines of equation~\whktifint, when combined with the
antisymmetry in $A,B$ of $g_{z^A\bar{w}^B}$, allow us to write
\eqn\mixedfromf{g_{z^A\bar{w}^B} = (\partial_{z^A} \partial_{\bar{w}^B}
- \partial_{z^B} \partial_{\bar{w}^A}) L
; \qquad \qquad
g_{w^A\bar{z}^B} = (\partial_{w^A} \partial_{\bar{z}^B}
- \partial_{w^B} \partial_{\bar{z}^A}) L
}
where $L$ is some real (by hermiticity of the metric---and
therefore identical in
the two equations~\mixedfromf) function.
Inserting equation~\mixedfromf\ into the third equation of~\whktifint\
gives
\eqn\partdiag{\partial_{\bar{w}^B} \bigl( g_{z^A\bar{z}^C} -
L,_{z^A\bar{z}^C} \bigr) - (B \leftrightarrow A) = 0,}
and therefore,
\eqn\nextpartdiag{g_{z^A \bar{z}^B} = L,_{z^A \bar{z}^B} +
\partial_{\bar{w}^A} G_{\bar{z}^B}}
for some integration one-form $G_{\bar{z}^B}$.  Combining this with the
fourth equation
of~\whktifint\ gives $G_{\bar{z}^B} = L,_{w^B}$.  Thus we have obtained
equation~\gol, which is the desired result.

We have shown that integrability of the quaternionic structure implies the
existence of
a potential $L$ for the metric.  Although equation~\gol\ holds only in a
coordinate system in which the quaternionic structures are constant,
equation~\tpr\ is coordinate invariant.  Equation~\tyi\ motivates us to ask
whether or not the existence of a potential $L$ obeying
equation~\tpr\ is generically implied by the weak HKT conditions,
independent of integrability of the quaternionic structure. 

\listrefs
\bye